\documentclass[copyright,creativecommons]{eptcs}

\providecommand{\event}{TFPiE 2022}
\providecommand{\thisvolume}[1]{This volume of EPTCS, Open Publ.\ Assoc}

\usepackage{amsmath}

\usepackage{url}
\usepackage{hyperref}
\usepackage{xcolor}
\usepackage{underscore}            

\usepackage[noabbrev,capitalise]{cleveref}

\usepackage{stfloats}

\usepackage[T1]{fontenc}
\usepackage[frozencache=true]{minted}
\setminted{autogobble,breaklines}
\newcommand{\hask}[1]{\mintinline{haskell}{#1}}

\usepackage{booktabs}
\usepackage{multirow}
\renewcommand{\arraystretch}{0.95}
\newlength{\tableunit}
\newcommand{\tableboxColored}[5]{%
    \multirow{#1}{*}{%
        \parbox[t][][t]{#5\columnwidth}{%
            \begin{tcolorbox}[size=fbox,height=#1\tableunit-1pt,%
                colback=#3,colframe=#4]
                \footnotesize
                \raggedright
                #2
            \end{tcolorbox}
        }
    }
}
\newcommand{\tableboxColoredWide}[4]{\tableboxColored{#1}{#2}{#3}{#4}{0.290}}
\newcommand{\tableboxColoredThin}[4]{\tableboxColored{#1}{#2}{#3}{#4}{0.158}}
\definecolor{darkblue}{rgb}{0.0, 0.2, 0.6}
\definecolor{darkgreen}{rgb}{0.0, 0.6, 0.2}
\definecolor{darkbg}{rgb}{0.0, 0.4, 0.4}
\newcommand{\tablebox}[2]{\tableboxColoredWide{#1}{#2}{black!5}{black!60}}
\newcommand{\tableboxB}[2]{%
	\tableboxColoredWide{#1}{#2}{darkblue!10}{darkblue!80}}
\newcommand{\tableboxG}[2]{%
	\tableboxColoredWide{#1}{#2}{darkgreen!10}{darkgreen!80}}
\newcommand{\tableboxGThin}[2]{%
	\tableboxColoredThin{#1}{#2}{darkgreen!10}{darkgreen!80}}
\newcommand{\tableboxBG}[2]{%
	\tableboxColoredWide{#1}{#2}{darkbg!10}{darkbg!80}}

\usepackage{tcolorbox}
\newenvironment{typedefn}
    {\begin{tcolorbox}[size=fbox]\hfill}
    {\hfill\null\end{tcolorbox}}
\newenvironment{exercise}
    {\begin{tcolorbox}[size=fbox]\hfill}
    {\hfill\null\end{tcolorbox}}
\usepackage{multicol}

\usepackage{tikz}
\usetikzlibrary{decorations.pathreplacing,calligraphy}

\newcommand{\class}{\mathcal}
\newcommand{\property}{}
\newcommand{\tool}{\textsc}
\newcommand{\foreign}{\textit}
\newcommand{\ma}{\mathtt{a}}
\newcommand{\mb}{\mathtt{b}}
\renewcommand{\epsilon}{\varepsilon}
\renewcommand{\phi}{\varphi}

\title{
    Teaching Simple Constructive Proofs with Haskell Programs
}
\author{Matthew Farrugia-Roberts
\institute{
The University of Melbourne\\
Melbourne, Australia}
\email{matthew@far.in.net}
\and
Bryn Jeffries
\institute{Grok Academy\\
Sydney, Australia}
\email{bryn.jeffries@grokacademy.org}
\and
Harald S{\o}ndergaard
\institute{
The University of Melbourne\\
Melbourne, Australia}
\email{harald@unimelb.edu.au}
}
\def\titlerunning{Teaching Simple Constructive Proofs with Haskell Programs}
\def\authorrunning{M.\ Farrugia-Roberts, B.\ Jeffries, \& H.\ S{\o}ndergaard}
\begin{document}
\maketitle

\begin{abstract}
    In recent years we have explored using Haskell alongside a traditional
    mathematical formalism in our large-enrolment university course on topics
    including logic and formal languages, aiming to offer our students a
    programming perspective on these mathematical topics.  
    We have found it possible to offer almost all formative and summative 
    assessment through an interactive learning platform, using Haskell as a 
    \foreign{lingua franca} for digital exercises across our broad syllabus.
    One of the hardest exercises to convert into this format are traditional
    written proofs conveying constructive arguments.
    In this paper we reflect on the digitisation of this kind of exercise.
    We share many examples of Haskell exercises designed to target similar
    skills to written proof exercises across topics in propositional logic
    and formal languages, discussing various aspects of the design of such
    exercises.
    We also catalogue a sample of student responses to such exercises.
    This discussion contributes to our broader exploration of programming
    problems as a flexible digital medium for learning and assessment.
\end{abstract}

%

\section{Introduction}
\label{sec:intro}

We teach \textit{Models of Computation}, a large-enrolment university course
on introductory topics in logic, discrete mathematics, formal languages, and
computability \cite{handbook}.
Most of our students take the course as part of their degree in computer
science or software engineering, wherein they learn how to use code to build
complex software systems. 
In our course, an important goal is for our students to learn the language of
mathematical modelling and proof, to help them think and communicate with
precision and rigour.

Over several years, we have experimented with a `programming approach'
to learning and assessment activities~\cite{Farrugia+2022},
via the Haskell programming language~\cite{haskell}
and the web-based programming education platform Grok Academy~\cite{grok}.
We use Haskell as a stepping stone to traditional mathematical formalism, 
offering students a programming perspective that leverages their computing
background. Haskell also serves as a \foreign{lingua franca} for our students
to express themselves in exercises spanning our rather broad syllabus in a
unified digital learning interface.
We have found Haskell problems to be a flexible medium for exercises in
many topics beyond programming skill itself.

Traditional exercises of one important class---written proofs---have been
difficult to digitise. To this end, we have designed a class of Haskell
exercises targeting some of the same skills as certain simple constructive
proof exercises. 
In particular, our students implement the central \emph{construction
algorithm} of a constructive proof as a Haskell function.
This paper details our approach to partially digitising constructive proof
exercises.
We contribute:
(1)~a large catalogue of example \emph{proof-style Haskell exercises} from
    our course, covering topics in propositional logic and formal
    languages (\cref{sec:examples});
(2)~an analysis of student responses to two questions in a recent exam
    (\cref{sec:analysis}); and
(3)~a discussion of similarities and differences between written proof
    exercises and our proof-style Haskell exercises (\cref{sec:discussion}).

This work also contributes to our broader initiative of digitising, or
`programmifying', our curriculum with Haskell and Grok Academy, which
we have outlined in recent work~\cite{Farrugia+2022}.
The current paper demonstrates in detail how the expressiveness of
a programming language affords the design of rich, open-ended digital
exercises for non-programming learning goals.

\section{Course description}
\label{sec:background}

\begin{table}[t!]
    \centering
    \begin{tabular}{@{}cccc@{}}
    \toprule
    Week
        & Topic in lectures and tutorials
        & \multicolumn{2}{c}{Haskell exercises for learning and assessment}
    \\ \midrule
    1
        & \tablebox{1}{Introduction}
        & \tableboxG{4}{
            Introduction to basic Haskell
            (self-paced, self-contained tutorial on
            functions, recursion, lists,
            basic algebraic data types)
        }
        & \\
    2
        & \tablebox{4}{
            Logic
            (syntax and semantics of
            propositional and predicate
            logic, and mechanised reasoning via
            resolution algorithms)
        }
        & & \\
    3   & & & \\
    4   & &
        & \tableboxGThin{8}{
            Worksheets: 6\% \\
            (four fortnightly formative tasks on algorithms for
            propositional logic, regular languages, and formal grammars)
        } \\
    5   &
        & \tableboxBG{3}{
            Assignment 1: 12\%\\
            (mathematical and algorithmic challenges in logic)
        }
        & \\
    6
        & \tablebox{2}{
            Discrete mathematics (sets, functions, relations, termination)
        }
        & & \\
    7   & & & \\
    8
        & \tablebox{4}{
            Formal languages
            (finite automata, regular expressions, context-free
            grammars, pushdown automata, Turing machines)
        }
        & \tableboxBG{4}{
            Assignment 2: 12\%\\
            (mathematical and algorithmic
            challenges in discrete mathematics
            and formal languages)
        }
        &
        \\
    9   & & & \\
    10  & & & \\
    11  & & & \\
    12  & \tablebox{1}{
            Computability (undecidability)
        }
        & & \\
    Exams 
        & & \tableboxB{1}{Exam: 70\% (3 hours, all topics)} & \\
    \bottomrule
    \end{tabular}
    \caption{
        \textit{COMP30026 Models of Computation}, 
        example semester calendar. Adapted from \cite{Farrugia+2022}.
	\label{tab:calendar}
    }
\end{table}

We begin by describing our course and our students.
Our course is an introduction to logic, discrete mathematics, formal
languages, and computation. \Cref{tab:calendar} gives an overview of the
topics studied. The emphasis of the intended learning outcomes is in
(1) \emph{applying} topics in logic and discrete mathematics to reason
about computational problems, and 
(2) \emph{analysing} and \emph{creating} computational models (from
finite-state automata to Turing machines) \cite{handbook}.

In particular, we consider it a learning goal for the students to be able to
apply their understanding of first-order logic and to analyse computational
models by carrying out simple proofs regarding these models. For example,
a student should be able to prove simple properties of formal language
classes, show by construction that a given language belongs to a given class,
or prove that a given language is outside a class by applying a pumping
lemma or a reduction argument.
Though the emphasis is on applications in computation we also study proofs
in other areas such as
propositional and predicate logic.

The course has a large enrolment, 
with over 500 students in the 2021 offering.
Most of our students take the course for either their 
undergraduate major or their coursework graduate program, 
in either computer science or software
engineering.
Our students are familiar with one or more 
programming languages as a prerequisite. 
Haskell is \emph{not} a prerequisite, though some students
take a concurrent elective in functional programming.
Most of our students have some university-level mathematics experience, 
but the course is considered mathematically demanding by many of our
students---especially when it comes to learning to write proofs.

\section{Learning with Haskell and Grok Academy}
\label{sec:programming-to-learn}

We have adopted Haskell exercises in learning and assessment tasks.
The exercises are delivered through a state-of-the-art web-based 
programming education platform, Grok Academy (Grok) \cite{grok}.
In 2021 we retained a small number of pen-and-paper exercises in assignments,
primarily for exercising and assessing students' proof-writing skills.
Haskell exercises comprise all other assessment, including the final exam.
We give an overview of this approach in recent work~\cite{Farrugia+2022},
and summarise the key elements here.

Since Haskell is not a prerequisite, our semester begins with a self-paced
introduction to a minimal subset of features we use.
Grok is a suitable platform for this introduction---designed for learning to
program, Grok offers a problem-based e-textbook interface.
Moreover, Grok is familiar to many of our students who have already used it 
in their introductory programming courses (Python, C, and/or Java).

However, programming is not one of our direct learning goals. 
Our students' main use of Grok is for learning \emph{with} Haskell, not
learning Haskell.
We offer formative and summative assessment through \emph{repurposed}
programming problems, specially designed to target our mathematical learning
goals.

\begin{figure}[ht!]
\centering
\begin{tikzpicture}
    \node[anchor=south west,inner sep=0,draw,thick] (screenshot) at (0,0)
      {\includegraphics[width=0.85\textwidth]{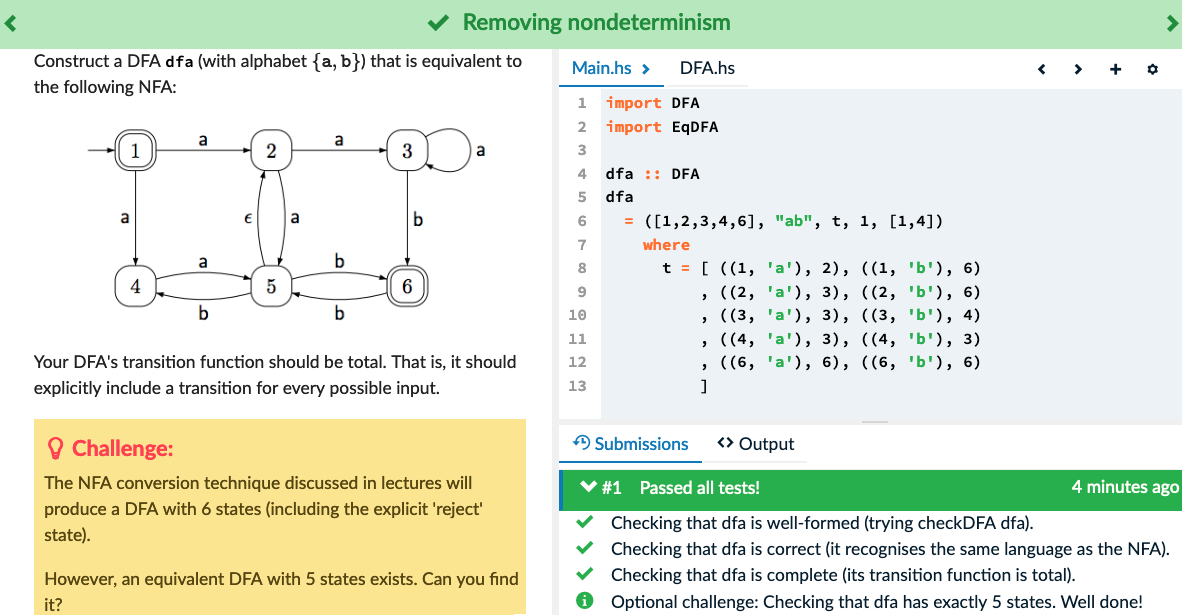}};
    \begin{scope}[x={(screenshot.south east)},y={(screenshot.north west)}]
        \begin{scope}[every path/.style={draw,thick,decorate,decoration={%
                calligraphic brace,raise=1em,amplitude=1em}}]
            \path (0,0.01) -- node[xshift=-3.1em] {(a)} (0,0.99);
            \path (1,0.91) -- node[xshift=+3.1em] {(b)} (1,0.32);
            \path (1,0.30) -- node[xshift=+3.1em] {(c)} (1,0.00);
        \end{scope}
    \end{scope}
\end{tikzpicture}
\caption{
    From the student perspective, a programming problem comprises
    (a) a description (shown: an embedded diagram);
    (b) a Haskell program editor;
    (c) feedback messages from a suite of tests.
    From~\cite{Farrugia+2022}.
\label{fig:grok}
}
\end{figure}

These exercises typically centre around a \emph{representation}:
a Haskell type capturing some mathematical object, such as a propositional
logic formula, a relation, or a finite-state automaton.
With these representations, we offer students two main kinds of programming
problems:
\begin{itemize}
    \item \emph{Instance problems}, where students answer a traditional
        short-answer-style exercise by defining an instance of the
        representation type. For example, we may ask students to design a
        regular expression for the language of odd-length binary strings, and
        expect them to respond with a Haskell representation of the
        expression $(0\cup1)((0\cup1)(0\cup1))^\ast$.
    \item \emph{Implementation problems},
        where students implement a topical algorithm involving the relevant
        representations, such as a resolution algorithm for propositional
        formulas in conjunctive normal form,
        or an algorithm for determinising a non-deterministic finite
        automaton.
\end{itemize}
To configure each problem in Grok, we supply 
(1)~a problem description,
(2)~skeleton code (including the representation definition, 
scaffolding, and supporting libraries), and 
(3)~a suite of test cases and associated feedback messages.
A student reads the description and writes their answer in 
Grok's web-based programming interface. 
They may run the test cases on their answer at any point. 
Depending on the configuration, the results and/or contents 
of the test may or may not be revealed to the student.
\Cref{fig:grok} shows the student interface in Grok, 
with an example \emph{instance problem} using a representation 
for deterministic finite automata (see \cref{sec:reglang}).

The definition of a `test case' is very flexible, allowing essentially
arbitrary Haskell programs to analyse the student's answer.
As a result in many cases, especially for instance problems, we can
\emph{verify}, rather than merely `test', a student's response, and
sometimes we can also provide helpful contextual feedback.
For example, in the automaton exercise in \cref{fig:grok}, 
or the regular expression exercise mentioned above,
we can run one test to determine if the language of the student's answer
precisely matches the desired language. If not, we can provide the student
with example strings their solution misclassifies.

We think this approach has the potential to lead to many pedagogical and
logistic benefits, provided our students can overcome the concomitant
language-related barriers~\cite{Farrugia+2022}.
Realising these benefits requires the careful design of Haskell exercises.
In this paper, we expand upon our recent overview \cite{Farrugia+2022}
with a deep focus on the design of one kind of Haskell exercise.
These exercises aim to address the challenging learning goal of
\emph{proof-writing} by partially mimicking a traditional proof exercise.

\section{Related work}
\label{sec:related}

There is a rich tradition for engaging teaching of functional programming,
with great ideas for exciting projects and exercises.
More recently, the challenges of effective feedback provision for
formative assessment, the quest for rapid marking for
summative assessment in very large classes, and a worldwide pandemic
have spurred developments in reliable auto-marking~\cite{Paiva22}.
This includes the marking of
functional programming tasks~\cite{praktomat16,KappelmannTFPIE22}.

It has long been realised that a functional programming language 
can serve as a powerful learning aid for topics \emph{beyond} programming.
Perhaps the most obvious such topic is discrete mathematics.
Several authors have pointed to the value that lies, for a student,
in the ability to implement and experiment with discrete structures,
based on \emph{executable} definitions written in a formalism that is
close to familiar mathematical 
notation~\cite{HendersonTFPiE2002,ODonnellHallPage2006,VanDrunen2017}.
Some authors extend this to topics in logic~\cite{Doets2004},
and there is a body of work on the teaching of proof techniques firmly 
grounded in declarative programming (e.g.,~\cite{hendriks2010,osera2013}).
Moreover, functional programming is sometimes introduced following a
``correct-by-construction'' philosophy, that is, it is taught 
hand in hand with program verification~\cite{KappelmannTFPIE22}.
One tool for this is \tool{Cyp} (``check your proof'') \cite{cyp13,cyp22},
with support for auto-marking of proof exercises.

Some use functional programming to teach finite-state machines and
related topics.
For example, Stoughton~\cite{StoughtonFPDE08} has used a domain-specific 
language (DSL) embedded in Standard ML, 
for the learning of formal language theory.
Similarly, \tool{FSM}~\cite{fsm2014} offers a DSL embedded in a Lisp
variant, for work with state machines.
Independently of functional programming,
the teaching and learning of automata theory and formal languages 
have been helped greatly by a range of sophisticated interactive 
graphical learning tools,
of which \tool{JFLAP}~\cite{jflap,Rodger:SIGCSE06} is a prominent example.

The aims of all these tools resemble ours, 
but we use Haskell \emph{uniformly across our varied syllabus}, on a
state-of-the-art interactive learning platform, that (most of) our students
already know from their programming education.
The use of a functional programming language as a \foreign{lingua franca}
for formative and summative assessment across such a wide syllabus
(including constructive proof exercises) seems rare.
The closest approach we know of is Waldmann's use of
\tool{autotool}~\cite{autotool}.
\tool{autotool} has a simple web-based interface and offers Haskell 
exercises in non-programming topics such as logic~\cite{autotool2014},
computation~\cite{autotool2002}, 
and others~\cite{autotool2014,autotool2015de,autotool2017de,autotool2017}.
Those exercises include implementation tasks and short-answer questions 
using Haskell as an expression language.
Rahn and Waldmann notably demonstrated a Haskell-function-based pumping
lemma exercise~\cite{autotool2002}, following a similar analogy to our own. 

The use of proof assistants 
(including those based on functional programming)
is orthogonal to our programming-based learning approach.
Such tools could potentially fill the gaps we leave through our 
`construction only' approach to proof.
But we note that the teaching of proof techniques is very different from 
the application of proof assistants.
In our context, where students are trying to learn proof principles, 
a proof assistant
may often \emph{hide} proof details that we would want to expose, 
such as the details of an inductive step.
Examples of proof assistants used in teaching logic are common
\cite{hendriks2010,nipkow2012,osera2013}.
We are not aware of examples in introductory formal languages, although
there are undoubtedly such cases.

\section{Constructive proof exercises and construction problems}
\label{sec:analogy}

In general, we have found proof exercises challenging to digitise as Haskell
exercises.
However, the algorithmic aspects of certain constructive proofs lend
themselves to digitisation.
In this section, we lay out the partial analogy we have drawn between
written constructive proof exercises and a certain kind of Haskell exercise
we call \emph{construction problems}.

Many of the important results in our syllabus are of a simple logical form:
``for all objects in some class $\class{X}$, there exists an object in some
class $\class{Y}$ such that the object has some property $\property{P}$''.
Examples include proofs of closure properties of formal language classes,
or that a certain restricted logical form is universal.
Moreover, the most appropriate method of proof is often a direct constructive
argument that
(1) demonstrates how to build an object in class $\class{Y}$ 
    systematically from the object in class $\class{X}$, and
(2) justifies that the constructed object indeed has the property
    $\property{P}$.

Our analogy captures the systematic construction (1) as a Haskell function.
Given a data type \hask{X} representing objects of class $\class{X}$,
and a data type \hask{Y} representing objects of class $\class{Y}$, 
the systematic construction can be implemented as a Haskell 
function of type \hask{X -> Y}.
A \emph{construction problem} is then a Haskell exercise that asks students to
implement such a function, framed as a way of proving the original result.
Construction problems are thus a kind of \emph{implementation problem} in
the terminology of \cref{sec:programming-to-learn}.

We contend that construction problems evoke similar skills to writing
constructive argument at the heart of a pen-and-paper proof, with Haskell
serving as an alternative mathematical notation.
First, implementing the construction algorithm requires a working 
understanding of the `input' (class $\class{X}$) and `output' 
(class $\class{Y}$) objects.
In particular, it requires enough understanding to manipulate their 
Haskell representations.
Moreover, the implementation requires a rigorous understanding 
of the details of the construction algorithm. 
Similar rigour is required as for a precise written description of the
algorithm.

We emphasise that the analogy is \emph{partial}. The Haskell function mirrors
the written construction part of the written proof, not the entire written
proof.
In particular, construction problems do not require the students to justify
that the constructed objects indeed have the property~$\property{P}$.

Nevertheless, this partial analogy suggests that at least \emph{some} of what
is pedagogically valuable in written proof exercises may also exist in
construction problems.
We return to discuss important learning goals missed by construction
problems, and other related topics, in \cref{sec:discussion}.
For now, let us take a tour through our syllabus, looking for constructive
proof exercises to convert into construction problems.

\section{Example construction problems from our course}
\label{sec:examples}

Here we focus on the design of construction problems for our course,
exemplifying the analogy outlined in \cref{sec:analogy}.
We list the representations and constructions relevant to propositional
logic (\cref{sec:proplogic}), regular languages (\cref{sec:reglang}), 
and non-regular languages (\cref{sec:contextfree}), 
and we showcase several example Haskell exercises in some detail.
Throughout, we remark on the design of construction problems based on our
experience, and how the exercises fit into our broader Haskell-based
approach.
Finally, in \cref{sec:other}, we comment on several examples 
using slight variations of the analogy.

\subsection{Examples from propositional logic}
\label{sec:proplogic}

Our students study propositional logic primarily as a tool 
for modelling and analysing computational problems.
We use a Haskell data type to represent propositional expressions,
as outlined in \cref{type:Exp}.

\begin{figure}[ht!]
\begin{typedefn}
\begin{minipage}[t]{0.21\textwidth}
    \begin{minted}{haskell}
        data Exp
          = VAR  Char
          | NOT  Exp
          | AND  Exp Exp
          | OR   Exp Exp
          | IMPL Exp Exp
          | BIIM Exp Exp
          | XOR  Exp Exp
    \end{minted}
\end{minipage}
\quad
\begin{minipage}[t]{0.40\textwidth}
    \paragraph{Example:} The formula:
    $$X \land (X \Rightarrow Y)$$
    would be expressed in Haskell with the following code:
    \medskip
    \begin{minted}{haskell}
        AND (VAR 'X')
            (IMPL (VAR 'X') (VAR 'Y'))
    \end{minted}
\end{minipage}
\quad
\begin{minipage}[t]{0.31\textwidth}
    \paragraph{Variants:}
    In some cases, we include \hask{FALSE} and \hask{TRUE} constructors for
    propositional constants, and/or allow \hask{VAR} labels of types other
    than \hask{Char}.

    It would also be possible to define the type with infix constructors.
\end{minipage}
\end{typedefn}
\caption{The \hask{Exp} type, our representation for propositional logic
    expressions.
\label{type:Exp}
    }
\end{figure}

This representation captures the recursive structure of propositional
expressions. This structure is convenient when defining recursive
constructions.
For instance problems, the syntax quickly becomes cumbersome---hence we also
provide a string parser utility so that students can write answers such as
\hask{parseExp "X & (X => Y)"}.

For the testing/verification of exercises involving propositional expressions,
we use Haskell implementations of
Wang's algorithm~\cite{Wang1960} for checking propositional entailment or
    equivalence, and
the Tseitin transform \cite{Tseitin68} for efficiently encoding expressions
    in 3-CNF (for processing by a third-party SAT solver).
Though testing propositional formulas is NP-complete in general, students
typically face small instances. We find that we can comfortably provide
automatic feedback within seconds.

A neat feature of our approach is that students get to implement their own
version of these tools in early worksheets, giving them at once tools to
support and test their own work, and also a deeper understanding of the
tools we use in testing.

\subsubsection{Functional completeness constructions}

As part of the study of propositional logic, we explore the 
expressiveness of certain logical connectives. 
In particular, we explore the diverse \emph{functionally complete} 
combinations of connectives (cf.\ Post's results~\cite{Post:func-compl41}). 
To prove a set of connectives functionally complete, it suffices to show how
every formula expressed with a set of connectives known to be complete can be
equivalently expressed with the set in question.
That constructive argument is readily cast as a construction problem
(where the student implements the translation into the restricted form).
We outline an example in \cref{fig:functional-completeness}.
We revisit this example, denoted FC, in \cref{sec:analysis}.

\begin{figure}[ht!]
\begin{exercise}
    \begin{minipage}[t]{0.30\textwidth}
        \paragraph{Exercise description:}
        Prove that the set $\{\Rightarrow, \oplus\}$ of connectives is
        functionally complete. Do this by writing a function
        \hask{tr :: Exp -> Exp} that translates arbitrary propositional
        formulas into equivalent formulas using no other connectives.

        \paragraph{Testing:}
        Check that output
        uses $\Rightarrow$ and $\oplus$ only
        and is equivalent to input.
    \end{minipage}
    \quad
    \begin{minipage}[t]{0.66\textwidth}
        \paragraph{Haskell answer:} A recursive construction.
        \begin{minted}{haskell}
            tr :: Exp -> Exp
            tr (VAR x)    = VAR x
            tr (IMPL e f) = IMPL (tr e) (tr f)
            tr (XOR e f)  = XOR (tr e) (tr f)
            tr FALSE      = XOR (VAR 'P') (VAR 'P')
            tr TRUE       = IMPL (VAR 'P') (VAR 'P')
            tr (NOT e)    = IMPL (tr e) (tr FALSE)
            tr (AND e f)  = IMPL (IMPL (tr e) (IMPL (tr f)
                                         (tr FALSE))) (tr FALSE)
            tr (OR e f)   = IMPL (IMPL (tr e) (tr FALSE)) (tr f)
            tr (BIIM e f) = IMPL (XOR (tr e) (tr f)) (tr FALSE)
        \end{minted}
    \end{minipage}
\end{exercise}
\caption{%
    An example propositional logic construction problem for a functional
    completeness result.
    We return to this example, denoted FC, in \cref{sec:analysis}.
}
\label{fig:functional-completeness}
\end{figure}

\subsubsection{Normal form constructions}

A similar class of exercises arises from studying various \emph{normal
forms}. Well-known examples include conjunctive normal form, disjunctive
normal form, negation normal form, and XOR normal form.
We are also free to define our own normal forms
(as in FC above).
A constructive proof that some set of formulas is a normal form involves
translating arbitrary formulas to equivalent formulas in the set.
Implementing this translation constitutes a constructive problem.

\subsection{Examples from regular languages}
\label{sec:reglang}

A rich vein of constructive results comes from the theory of regular
languages, since they are defined by the existence of some simple automaton
or expression, which must usually be constructed.

\begin{figure}[ht!]
\begin{typedefn}
    \begin{minipage}[t]{0.47\textwidth}
        \begin{minted}{haskell}
            type State  = Int
            type Symbol = Char
            type DFA
              = ( [State]       -- states
                , [Symbol]      -- alphabet
                , [((State, Symbol), State)]
                                -- transitions
                , State         -- start state
                , [State]       -- accept states
                )
        \end{minted}
        \vspace{-1.5\baselineskip}
        \paragraph{Variants:}
        We sometimes use other state types.
        
        We represent the transition function as an explicit relation.
        One could use a Haskell function of type
        \hask{State -> Symbol -> State}.
    \end{minipage}
    \quad
    \begin{minipage}[t]{0.47\textwidth}
        \paragraph{Example:}
        The DFA traditionally depicted:
        \medskip
        \begin{center}
        \includegraphics[width=0.5\textwidth]{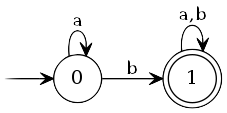}
        \end{center}
        is then expressed with the Haskell code:
        \begin{minted}{haskell}
            ([0,1], "ab", d, 0, [1])
              where
                d = [ ((0, 'a'), 0)
                    , ((0, 'b'), 1)
                    , ((1, 'a'), 1)
                    , ((1, 'b'), 1)
                    ]
        \end{minted}
    \end{minipage}
\end{typedefn}
\caption{%
    The \hask{DFA} type, our representation for deterministic finite
    automata.
    \label{type:DFA}
}
\end{figure}

We offer Haskell data types for deterministic finite automata (DFAs),
non-deterministic finite automata (NFAs), and regular expressions. 
The DFA representation (\cref{type:DFA}) directly mirrors the standard
mathematical `5-tuple' definition.
The NFA type is similar, but with multiple start
states\footnotemark.
Regular expressions use a recursive type (\cref{type:RegExp}). Like for
propositional formulas, the type is optimised for expressing recursive
algorithms, and we provide a string parser for expressing instances.
\footnotetext{Our Haskell and mathematical NFAs allow a \emph{set} 
of start states, as needed in Brzozowski's minimisation
method~\cite{Brzozowski:minimisation62,Watson:phd95}.}

\begin{figure}[ht!]
\begin{typedefn}
    \begin{minipage}[t]{0.33\textwidth}
        \begin{minted}{haskell}
            data RegExp
              = Symbol Char
              | EmptyStr
              | EmptySet
              | Concat RegExp RegExp
              | Union  RegExp RegExp
              | Star   RegExp
        \end{minted}
    \end{minipage}
    \quad
    \begin{minipage}[t]{0.63\textwidth}
        \paragraph{Example:}
        The regular expression
            $$0\ast(1 \cup \varepsilon)$$
        could be expressed in Haskell with the following code:
        \medskip
        \begin{minted}{haskell}
            Concat (Star (Symbol '0'))
                   (Union (Symbol '1') EmptyStr)
        \end{minted}
    \end{minipage}
\end{typedefn}
    \caption{The \hask{RegExp} type, our representation for regular
    expressions
    \label{type:RegExp}
    }
\end{figure}

When testing exercises involving DFAs, NFAs, and regular expressions, we can
provide comprehensive analysis of instances using a suite of Haskell
tools for converting NFAs and regular expressions to DFAs (see
\cref{sec:examples:reglang:conversions}), and then testing equivalence
between the resulting DFA and a known solution DFA (by testing if the
symmetric difference automaton accepts any strings).
This means we can `test' the correctness of individual DFAs, NFAs, and regular
expressions with perfect accuracy, compared to, say, checking a small number
of input strings.

Moreover, we can search exhaustively for any strings accepted by the
symmetric difference automaton and provide such strings as misclassified
examples to the student, as a form of rich, contextual feedback (a similar
approach is found in other systems~\cite{autotool2002,D'Antoni+2015a}).

The computational complexity of conversion to DFAs and constructing the
symmetric difference automaton limit the size of automata and expressions
we can analyse (while maintaining responsive automatic feedback).
In practice, these limitations do not manifest at the sizes of DFAs and
regular expressions we consider but rule out NFAs with more than around 10
states.

\subsubsection{Closure property constructions}

Regular languages are closed under the regular language operations (union,
concatenation, and the Kleene star), as well as the operations of language
intersection, complement, reversal, and many others.
Proving each such closure property requires an argument that takes
automata or expressions characterising arbitrary regular input language(s), 
and constructs an automaton or expression characterising the output 
language of the operation.
Implementing such constructions makes a suitable Haskell construction problem.
Figure~\ref{fig:skip-dfa} shows an example, 
adapted from an assignment question.

\begin{figure}[ht!]
\begin{exercise}
      \begin{minipage}[t]{0.46\textwidth}
        \paragraph{Exercise description:}
        Consider the operation $\mathit{skip}$ defined by
        $$
          \mathit{skip}(L)
          = \{
                x y
            \mid
                x z y \in L,
                x \in \Sigma^\ast,
                y \in \Sigma^\ast,
                z \in \Sigma
            \}
        $$
        Informally, the language $\mathit{skip}(L)$ contains all strings 
        obtained from the strings of $L$ by removing any one symbol.
        For example:
        \begin{itemize}
        \item
        if $L = \{\epsilon,\ma\mb\}$, then $\mathit{skip}(L) = \{\ma,\mb\}$
        \item
        if $L = \{\ma,\mb\}$, then $\mathit{skip}(L) = \{\epsilon\}$
        \end{itemize}
        Show that the class of regular languages is closed under
        $\mathit{skip}$ by writing a Haskell function 
        \hask{skip :: DFA -> NFA} such that
        $$
          \mathcal{L}(\mbox{\hask{skip}}\ d)
          = \mathit{skip}(\mathcal{L}(d))\,.
        $$

        Hint: Build an NFA containing two layers with copies of the initial
          DFA. Think about how to connect the layers and how to define the
          start and accept states.
        \vspace{-.5\baselineskip}
        \paragraph{Testing:} For several inputs, check that output
          NFA is equivalent to known correct DFA.
      \end{minipage}
      \quad
      \begin{minipage}[t]{0.47\textwidth}
        \paragraph{Haskell answer:} Following the hint:
        \begin{minted}{haskell}
          skip :: DFA -> NFA
          skip d@(qs, xs, ts, q0, as)
            = ( qs++qs'
              , xs
              , ts++ts'++ts''
              , [q0]
              , as'
              )
            where 
              -- With helper fn. 'renameDFA',
              -- make 2nd layer with distinct
              -- states
              rename
                = (+(1+(maximum (map abs qs))))
              (qs', _, ts', _, as')
                = renameDFA rename d
              -- Epsilon transitions between
              -- layers implement the skip:
              ts''
                = [ ((q, epsilon), rename r) 
                  | ((q, x), r) <- ts
                  ]
        \end{minted}
    \end{minipage}
\end{exercise}
\caption{An example construction problem based on a closure property of
    regular languages.
}
\label{fig:skip-dfa}
\end{figure}

We note that the choice of language representation 
influences the difficulty of the construction.
For example, showing closure under union is trivial 
for regular expressions and straightforward for NFAs, 
but requires some care with a DFA representation.
Conversely, closure under complement is straightforward in a DFA
representation, but for regular expressions or even NFAs, 
we are unaware of a direct method of construction 
(not involving, essentially, conversion to a DFA representation).

When it comes to proving these properties for their own sake, 
one has the choice of representations. 
In the pedagogical setting, there is value in exploring each method. 
The choice of representation can be made based on
the desired topic or difficulty level.
The strategic choice of representations is also a point of interest
for our students. 
While we have not explored this direction in our course,
it would possible to allow students to choose a representation 
as part of their response to a construction problem 
(by using an appropriate algebraic data type).

\subsubsection{Language family constructions}

A favourite class of constructions is to give students a parameterised
family of languages and ask them to prove that all languages in the family
are regular. 
It is sufficient to construct a DFA (for example) for each parameter, 
generalising the traditional task of designing a DFA for a particular language.
This exercise is readily cast as a construction problem, where the student
writes a function to build the DFA based on the parameter.
\cref{fig:loop-dfa} shows an example (LA) adapted from a final exam.

\begin{figure}[ht!]
    \begin{exercise}
        \begin{minipage}[t]{0.48\textwidth}
            \paragraph{Exercise description:}
            Consider the singleton alphabet $\Sigma = \{\mathtt{a}\}$.
            Given a positive natural number $d$, we can define a language of
            strings on $\Sigma$:
            $$M_d = \big\{\ \mathtt{a}^{n} \ \big|\ n \geq 0,\ n
            \text{ is a multiple of } d \big\} \,.$$
            In other words, $M_d$ is the language of all strings of
            `$\mathtt{a}$'s whose length is a multiple of $d$.

            Prove that all languages of this form are regular:
            Write a function \hask{m :: Int -> DFA} so that `\hask{m} $d$'
            returns a DFA recognising $M_d$ (for $d > 0$).
        \end{minipage}
        \quad
        \begin{minipage}[t]{0.46\textwidth}
            \paragraph{Testing:} Check that the output DFA is equivalent to
            a known solution DFA.
            \paragraph{Haskell answer:} A direct construction, closely
            mimicking a pen-and-paper definition.
            \begin{minted}{haskell}
                m :: Int -> DFA
                m d = (qs, "a", ts, 0, [0])
                  where
                    qs = [0..d-1]
                    ts = [ ((i, 'a'), (i+1) `mod` d)
                         | i <- qs
                         ]
            \end{minted}
        \end{minipage}
    \end{exercise}
    \caption{An example DFA construction based on a simple parameterised
    family of languages.
    We return to this example, denoted LA, in \cref{sec:analysis}.
    \label{fig:loop-dfa}
    }
\end{figure}

Many variations of this example are possible.
A challenging construction exercise we used in a previous assignment
highlights one of the advantages of using a single medium for exercises
across our broad syllabus. We designed a family of languages parameterised
by propositional logic formulas.
Given a formula, the corresponding language consisted of strings encoding
satisfying truth assignments for the formula.
Through a recursive DFA construction algorithm, students were able to prove
that this family of languages is regular.
The recursive construction was also an opportunity for students to use,
as helper functions, constructive algorithms for product and complement DFAs,
built in a previous worksheet.

\subsubsection{Representation conversion constructions}
\label{sec:examples:reglang:conversions}

A staple of an introductory course on formal languages are the constructive
proofs showing that DFAs, NFAs, and regular expressions all capture the same
class of languages, since for every regular expression there is a
corresponding DFA, and so on. 
Some of these construction algorithms are well-suited as complex 
implementation exercises. 
In particular, we typically include an exercise to implement an NFA 
determinisation algorithm in a worksheet, and we have in the past included, 
in an assignment, a guided implementation of Brzozowski's algorithm for 
converting regular expressions directly to DFAs~\cite{Brzozowski64}.

These same tools form part of our testing apparatus for regular language
exercises, which first converts a student's NFAs or regular expressions
to DFAs, as discussed above.
Once again, students can take part in the building of tools that we (and
they) use to analyse their work.

These conversions are also useful as `helper functions' in other exercises.
For example, after implementing the NFA determinisation function, our
students are a short step away from implementing Brzozowski's double-reversal
method for DFA minimisation~\cite{Brzozowski:minimisation62,Watson:phd95}.

Another construction algorithm of interest is the `union-free decomposition'
of regular expressions.
For any regular language, there is always a representation of the 
language as a union of a finite number of regular expressions which, 
themselves, do not contain the union operation \cite{nagy04-regular}.
In a challenging construction problem, which we have used in a final exam,
we explain this representation to students, and then ask them to write the
conversion algorithm.
We detail this example in \cref{example:union-free}.

\begin{figure}[ht!]
\begin{exercise}
      \begin{minipage}[t]{0.44\textwidth}
        \paragraph{Exercise description:}
            A regular expression is union-free if it does not use
            the union operator. A union-free decomposition of a regular
            language $R$ is a finite list of union-free regular expressions
            $r_1, \ldots, r_k$ such that
            $R = \mathcal{L}(r_1) \cup \cdots \cup \mathcal{L}(r_k)\,.$
            Prove that every regular language has a union-free
            decomposition. Do this by writing a Haskell function
            \hask{uf :: RegExp -> [RegExp]}
            that takes a regular expression \hask{r} and produces a
            union-free decomposition of $\mathcal{L}($\hask{r}$)$.

        Hint: The following rules for regular expressions may be useful.
          \newcommand\Lang{\mathcal{L}}
          $$\begin{aligned}
              \Lang\big(  r (r_1 \cup \cdots \cup r_n) \big)
              =\ & \Lang\big(  r\,r_1 \cup \cdots \cup r\,r_n \big) \\
              \Lang\big(  (r_1 \cup \cdots \cup r_n) r \big)
              =\ & \Lang\big(  r_1\,r \cup \cdots \cup r_n\,r \big) \\
              \Lang\big(  (r_1 \cup \cdots \cup r_n)\ast \big)
              =\ & \Lang\big(  (r_1\ast\,\cdots\,r_n\ast)\ast\big)
          \end{aligned}$$
      \end{minipage}
      \quad
      \begin{minipage}[t]{0.50\textwidth}
        \paragraph{Haskell answer:} Following the hint:
        \begin{minted}{haskell}
            uf :: RegExp -> [RegExp]
            uf (EmptyStr)   = [EmptyStr]
            uf (EmptySet)   = [EmptySet]
            uf (Symbol x)   = [Symbol x]
            uf (Union r s)  = uf r ++ uf s
            uf (Concat r s)
              = [Concat a b | a <- uf r, b <- uf s]
            uf (Star r)
              = [Star (cat [Star s | s <- uf r])]
        \end{minted}
        \vspace{-\baselineskip}
        \paragraph{Note:}
            A helper function, equivalent to \hask{cat = foldr1 Concat}
            was provided to students with documentation.
        \vspace{-\baselineskip}
        \paragraph{Testing:} For various inputs, check that the output
          decomposition is equivalent to the input regular expression.
    \end{minipage}
\end{exercise}
\caption{
    An example construction problem based on a representation of
    regular languages.
    \label{example:union-free}
}
\end{figure}

\subsection{Examples from non-regular languages}
\label{sec:contextfree}

In the study of more complex languages, our students complete exercises
involving Haskell representations of context-free grammars (CFGs),
(deterministic) pushdown automata ((D)PDAs), Turing machines, and other
models of computation.
For brevity, we omit a detailed description of the representations
here---suffice it to say that they are similar to the DFA representation
shown in \cref{type:DFA}, again mirroring the standard tuple-based
mathematical representation of these objects.

With instances of these models, we quickly move into the realm of the
undecidable.
It is not possible in general to decide if a student's model recognises a
particular language.
We have a suite of emulation tools, and can fall back on string-based
testing and, if necessary, manual inspection, which is usually sufficient
(students do not often submit pathological or even large instances).

Most of our non-regular language exercises involve the creation or analysis
of specific instances
(in the case of analysis, we can often design the exercise so that
correctness is decidable).
However, we have designed a small number of construction problems, as we
outline below.

\subsubsection{Closure property constructions}

As with regular languages, one class of construction exercises comes from
closure properties of higher language classes. For example, students can
write the construction by which one can see that the classes of context-free
and decidable languages are closed under the regular operations, or
other operations such as reversal, and intersection with regular languages.

\subsubsection{Representation conversion constructions}

Though we are yet to deploy examples in this vein, there are also several
representation conversion results for higher language classes that may
be suitable for construction problems.
For example, one could consider conversions between CFGs and PDAs, or
various Turing-equivalent models, and converting CFGs to Chomsky normal
form.
These detailed constructions could be worth students' implementing as an
assignment challenge or in a guided worksheet.

One class of conversions that we have considered is to convert regular
models into CFGs, PDAs, and TMs. A student can prove that the language
classes associated with these models contain the regular languages through
such constructions. Most of these constructions are quite straightforward
(e.g., making no use of the PDA stack or TM tape),
but may still be worthwhile for students.

A more challenging variation is to apply a restriction on the construction.
For example, it is possible to construct a deterministic PDA for any regular
language (say, represented as a DFA) using only three PDA states 
(regardless of the DFA size) by utilising the stack.
We detail this example in \cref{exercise:pda3}.
Similar constrained constructions
are available for TMs and may be possible for other models. 
To prove these results, students can implement the constrained 
construction algorithms in Haskell.

\begin{figure}[t]
    \begin{exercise}
        \begin{minipage}[t]{0.43\textwidth}
            \paragraph{Exercise description:}
            Prove that any regular language can be recognised by a
            deterministic PDA (DPDA) with just three states.

            To do this, write a Haskell function,
            \hask{dfa2pda :: DFA -> DPDA} that converts an arbitrary
            DFA to an equivalent DPDA with exactly three
            states.

            Hint: While there is a limit on the size of the PDA's state
            machine, there is no limit on the size of the PDA's stack
            alphabet.
            
            \vspace{\baselineskip}
            \paragraph{Haskell answer:}
            From the start state, load the second DFA state onto the stack
            while consuming the first symbol. Transition to an accept or
            non-accept state in the DPDA depending on where you would be in
            the DFA.  Carry out the remaining transitions using
            the symbol on the stack to represent the state in the DFA,
            and the DPDA state to represent whether the DFA is ready to
            accept or reject.
        \end{minipage}
        \quad
        \begin{minipage}[t]{0.52\textwidth}
            \begin{minted}{haskell}
                dfa2pda :: DFA -> DPDA 
                dfa2pda (qs, as, ts, q0, fs)
                  = ([0, 1, 2], as, qs, ts', 0, fs')
                where
                  accepts q
                    = if (elem q fs)  then 2 else 1
                  fs'
                    = if (elem q0 fs) then [0,2] else [2]
                  ts'
                    =  [ ( (accepts b, a, b)
                         , (accepts b', b')
                         )
                       | ((b, a), b') <- ts
                       ] 
                    ++ [ ( (0, a, eps)
                         , (accepts b', b')
                         )
                       | ((b, a), b') <- ts
                       , b == q0
                       ]
            \end{minted}
        \end{minipage}
    \end{exercise}
    \caption{
        An example construction problem based on a non-regular model of
        computation.
        \label{exercise:pda3}
    }
\end{figure}

As with many construction problems, but particularly in this case, we must
take care to avoid an overload of low-level Haskell details.
It is important to design exercises with clean implementations.

\subsection{Variations of construction problems}
\label{sec:other}

So far we have discussed construction problems based on constructive
algorithms of the form \hask{X -> Y} for some representation types
\hask{X} and \hask{Y}. In this section we collect notes on several
variations on this theme.

\subsubsection{Instantiation as scaffolding}

To each construction problem correspond many simpler problems of the form
`for this \emph{given} object of class $\class{X}$, construct the
corresponding object of class $\class{Y}$'.
While the general construction problem is a kind of
implementation problem, that is, a Haskell exercise with answer type
\hask{X -> Y}, these simpler problems correspond to instance
problems---Haskell exercises with answer type \hask{Y}.

When designing an assignment or exam, 
we can often calibrate the level of difficulty,
as well as target a different set of learning outcomes,
by changing between instance problems and implementation problems.
Moreover, a useful design pattern is to run instance
problems and implementation problems together.
For example, in one multi-part exam question, students might be presented
with one or two examples of a construction in the form of instance problems
and then be asked in a final part to implement the general construction. 
The instances serve as a `warm-up' for the general algorithm, 
and also give students credit for time spent considering examples as 
they prepare to provide a general answer in the final part.

\subsubsection{Proof-based framing of instance problems}

There are many places in our course where we might use the term `constructive
proof' in a slightly different sense than above, not referring to a
particular general construction algorithm, but to the construction of a
witness to an existential claim (or a counterexample to a universal one).

By a similar analogy to that underlying construction problems 
(of the implementation type),
we can draw an analogy between these instance constructions and instance
problems.
The distinction is that between a Haskell function with parameters, and a
constant Haskell function.
We often frame an instance problem this way, as a kind of `proof',
to more closely relate our Haskell instance exercises
to the mathematical topics of our syllabus.
For example, rather than
describing an exercise involving CFG ambiguity as ``give an example string
for which this CFG permits at least two distinct parse trees'', we might say
``prove that this CFG is ambiguous by giving a string for which \ldots''
Many other examples of this form exist, across propositional logic,
regular languages, non-regular languages, and also predicate logic and
discrete mathematics.

Once again, this analogy is \emph{partial}, since students usually are
not required to carefully justify the correctness of their examples.  
However, we can often verify correctness with appropriate tests
if the problems are carefully designed so that correctness remains decidable.

\subsubsection{Proof-based framing of algorithm implementation problems}

In a sense that is relevant to the topics of our course, all of the
implementation problems we study, whether framed as constructive proofs
(such as the examples earlier in this section) or as algorithms (such as
a DFA minimisation algorithm, Wang's algorithm, the Tseitin transform,
or a propositional resolution algorithm), are a kind of existence proof that
certain problems are decidable (and inhabit a certain complexity class).

For example, while we usually frame the students' implementation of the
Tseitin transform in an early worksheet as the completion of a useful tool 
for computing with propositional logic representations, it is equally an 
existence proof that a (polynomial-time) algorithm exists to convert an
arbitrary propositional formula to 3-CNF. This particular algorithm also
serves as a (polynomial-time) reduction between the Boolean satisfiability
problem for arbitrary formulas, and the 3-SAT problem.
Thus it provides a nice opportunity to link several topics within our broad
course.

Again we note that the analogy is \emph{partial}, as students are not
normally asked to carefully justify their implementation's correctness 
(or complexity).

\subsubsection{Higher-order constructions}

In logical terms, the construction problems we have examined in detail above
correspond to a constructive proof of a universally quantified existence
result. Similarly, appropriately-framed instance problems correspond to a
constructive proof of a simple existence result. 

The possibility of extending this analogy is suggested: Could we make a
Haskell exercise out of the proof of a higher-order result? For example,
a result of the form
``for all $x \in \class{X}$,
there exists $y \in \class{Y}$, such that
for all $z \in \class{Z}$,
there exists $u \in \class{U}$, such that the property
$\property{P}(x, y, z, u)$ holds.''

For example, a proof that a language does not have the pumping property for
regular languages has exactly the above form.
Such a proof, together with the well-known pumping lemma for regular
languages, is often sufficient to show that a language is non-regular.
The same holds for a proof that a language does not have the pumping property
for context-free languages.
Pumping lemma exercises are regarded by many of our students as a very
challenging part of our course, perhaps partly because of their mathematical
complexity.
If some Haskell-based exercise could assist the students in constructing a
pumping lemma proof, this could be pedagogically valuable.

It would be possible to ask students to implement functions that play the
role of the interacting agents in a query-based proof of the pumping property
for a language, for example.
This approach to the pumping property would be similar to the `pumping lemma
game' available in \tool{JFLAP}~\cite{jflap}. A similar, Haskell-based
approach has already been developed for \tool{autotool}~\cite{autotool2002}.

One concern is that the specification of the Haskell-based exercise could
retain the complexity of the mathematical approach, due to the necessary
use of higher-order functions and other challenging language features. If
this analogy is to be extended, it will not remove the inherent logical
complexity of the results to be studied.

\section{Student performance in high-stakes assessment}
\label{sec:analysis}

As we have gradually moved to our new assessment regime,
eventually delivering \emph{all} assessment through Grok,
we have naturally wondered what consequences
(predicted or unforeseen) there might be for students.
The concern was never that students might be uncomfortable
with online assignment submission and online exams---if anything,
there seems to be a growing percentage of students who feel
uncomfortable with pen-and-paper assessment and instead value the
better writing and editing features offered by online tools.
Instead our concern has been that we are forcing students to communicate
their knowledge in a new and unforgiving notation, namely Haskell.
There are arguable advantages and disadvantages involved,
for learning, feedback provision, learning support, 
student engagement, marking, and course management 
generally~\cite{Farrugia+2022}.
So how do students respond to the change?

We do not have data from surveys asking students directly
(apart from standard student experience questionnaires,
to which students continue to respond positively 
about the course generally).
Grok, however, provides us with logs that allow us to explore
students' activity, for example during exams.
This gives us some insight into whether
Haskell ends up being a significant barrier to students
expressing themselves, in particular in their answers to more
difficult problems, such as proof constructions.

In this section, we sample student responses to two exam questions. 
These are the questions we labelled FC and LA in Section~\ref{sec:examples}.
High-stakes assessment is arguably the most suitable context for
this analysis.
During an exam, students can submit answers to a problem repeatedly
until they are happy with their answer---only the latest submission counts.
In the exam setting they receive real-time feedback from the compiler,
and, where relevant, additional automated feedback about the
well-formedness of their answer, beyond syntax.
They do not see the results from any of our correctness tests.

\subsection[The FC example from Section 6.1]{The FC example from Section \ref{sec:proplogic}}
\label{sec:FC}

In the exam, 517 students submitted answers to this question, 
while 38 did not submit.
Students who did submit made 2.57 submissions, on average.
In the following, we analyse each student's \emph{final} answer only, 
in case they made multiple submissions.
Of the 517 answers, 483 passed the well-formedness tests and were
then marked for (partial) credit according to their degree of correctness.
The remaining 34 failed to compile.
According to \tool{ghci}'s feedback to the 34 students,
the errors were distributed as shown in Table~\ref{tab:errors}, 
second column.

\newcommand{\pad}{\hphantom{2}}
\begin{table}[t]
\begin{center}
\begin{tabular}{lcc}
   \hline
   \textbf{Issue}  &  
	~~ \textbf{FC (Fig.~\ref{fig:functional-completeness})} ~~ &
	~~ \textbf{LA (Fig.~\ref{fig:loop-dfa})} ~~ 
\\ \hline
   Indentation problem                       & \pad  3  &      10
\\ Syntax error                              & \pad  2  &      16 
\\ Variable scope error                      &      14  & \pad  7 
\\ Hole (`\verb!_!') used as expression      & \pad  1  & \pad  0  
\\ Nonlinear pattern                         & \pad  1  & \pad  0  
\\ Pattern overlap                           & \pad  0  & \pad  1  
\\ Bad enumeration                           & \pad  0  & \pad  2  
\\ Type error                                &      13  & \pad  6  
\\ \hline
   Generation of ill-formed DFA              & ---      &      22
\\ Failure to terminate                      &       0  & \pad  2 
\\ \hline
\end{tabular}
\end{center}
\caption{Number and kind of responses received by students regarding 
    ill-formed submissions.
    \label{tab:errors}
}
\end{table}

More careful scrutiny reveals that almost all (12 of 13) type errors
were the result of incorrect use of parentheses in function applications.
In essence, 12 students wrote something of form \verb!a b c! when they
meant \verb!a (b c)!.
Many scope errors came from code such as
\begin{minipage}{0.1\textwidth}
~~~~~~
\end{minipage}
\begin{minted}{haskell}
    tr FALSE = tr (AND (VAR x) (NOT (VAR x)))
\end{minted}
which shows a fair understanding of the propositional logic problem.
But the variable scope error prevents such
understanding from being recognised, even by sophisticated auto-marking.
For this reason, we mark all non-compiling submissions manually.
That way (possibly partial) marks can still be awarded for plausible code.
There were three cases of \tool{ghci} flagging an indentation error;
in all cases, these were caused by students using a layout similar to that
in Figure~\ref{fig:functional-completeness}, but leaving one or more cases 
unfinished, with no text after a defining `\verb!=!'.

To summarise, student performance on this question does not suggest
that Haskell has ``gotten in the way'', except for a small number of
students who struggle with the role of parentheses in a language
where simple juxtaposition is used for function application.

\subsection[The LA example from Section 6.2]{The LA example from Section \ref{sec:reglang}}
\label{sec:LA}

In the exam, 499 students submitted answers to this question,
while 56 did not submit.
Students who did submit made 2.39 submissions, on average.
Again the analysis we perform of the submitted answers involves 
only each student's final answer.
Of the 499 answers, 42 failed to compile.
\tool{ghci} feedback suggests the errors were distributed as listed 
in the third column of Table~\ref{tab:errors}.

Altogether 66 students would have received immediate feedback to warn
them that something about their answer was wrong.
Namely, in addition to the 42 with compiler errors,
two were warned about termination issues, 
and another 22 failed to generate well-formed DFAs.
In the last case, a message, auto-generated by us, explained in
what way the DFA was ill-formed (perhaps it was not deterministic, 
or it used letters outside its alphabet, or states outside its state set).
For 22 students, it appears this was not sufficient help, 
or alternatively, some students had insufficient time left
to fix the issue.

As the feedback generated at submission time was based on a single
test case ($d=5$), there may be cases of ill-formedness
(and non-termination) that go undetected.
Indeed, the full set of test cases identified another 11 cases of
ill-formed DFAs being generated---almost always for the corner case $d=1$.

The 10 indentation errors are mostly because of incorrectly placed
\hask{where} clauses.
Of the 16 syntax errors, 10 show difficulty with list comprehension syntax,
like pointing arrows in generators the wrong way.
The scope errors are all from expressions involving list comprehension.
This kind of error, where a student uses a variable outside its scope,
should not necessarily be taken as evidence that Haskell has hindered 
expression;
rather these cases mirror the identical mistake we see when
students use mathematical notation, such as set comprehension,
leaving variables free.

We mentioned that 56 students did not submit an answer. 
In some cases, the distinction between ``no attempt'' 
and ``syntax error'' is not clear.
For example, when a student submits
\begin{minted}{haskell}
    makeDFA d = (
\end{minted}
we classify that as a syntax error, but in all likelihood, it is
the result of a student deciding not to attempt the question
(and submitting the snippet nevertheless).
Of the 16 syntax error cases mentioned in Table~\ref{tab:errors},
third column,
five are arguably in this category, that is, they do not really
suggest a lack of knowledge of Haskell syntax.
Likewise, three of the 22 ill-formed DFAs 
come about because a student simply submits the bare code that was
provided as scaffolding for the question---evidence that
the student simply decided to give up this question.

Hence the number of student submissions 
(in the case of this particular question) 
that point to a struggle with Haskell notation is perhaps more accurately
given as 34, but in any case, it is in the vicinity of 7\% of submissions.
The number of students who appear to give up the question
(whether for lack of time or ability) is 64, that is, 
around 12\%---not unlike what we would previously see in a
typical pen-and-paper question of this kind.

To summarise, submitted answers for this question suggest that 
5--10\% of our students begin to struggle with Haskell once a
question calls for a combination of language features including
recursive definition, \hask{where} clauses and list comprehension.
To preserve fairness, manual marking seems unavoidable for answers 
that get rejected by \tool{ghci}.

\section{Discussion}
\label{sec:discussion}

We turn to discuss the various pedagogical and logistic dimensions of the
choice between written proof exercises and our proof-inspired Haskell
implementation exercises.

\subsection{Pedagogical evaluation}

In terms of the pedagogical value of these two learning activities, a
comparison must account for the skills which each task will exercise for
students, and the alignment of these skills with the intended learning
outcomes.
\Cref{sec:discussion:missing} discusses aligned skills neglected by
our Haskell exercises, and \cref{sec:discussion:barriers} discusses
additional skill requirements introduced by the change in exercise format.

\subsubsection{Additional skills exercised by written proofs}
\label{sec:discussion:missing}

As we have outlined above, we believe that well-designed Haskell
implementation exercises can bring out many of the same kinds of skills
as the writing of a detailed written description of a construction algorithm.
These include the ability to specify the construction itself with a high
level of precision, drawing on a detailed understanding of the formal objects
involved, and of the analysis and creation techniques embodied by the
construction. 
These all may be considered intended learning outcomes in many
maths-based classes such as our own.

As we produce a partial analogy, it is clear that the skills involved in
analysing and justifying one's construction are not exercised directly in the
same manner as in a written proof. If this is considered an important
learning outcome, it may be necessary to augment Haskell-based activities
with opportunities for students to practice and receive feedback on written
justifications, as we do in our course.

It is worth considering the role that rapid corrective feedback from
automated tests could play as a kind of substitute for explicit justification
tasks.
We suppose a student has an \emph{implicit} justification in mind guiding
the design of their constructive algorithm.
When they receive feedback from tests,
for example that their construction fails in some case, they may adapt their
implicit justification to resolve an error in their approach.
The student is exercising their justification skills in a minor way, 
though not the skill of formalising and explicating their justification.
On the other hand, unlimited rapid feedback may lead
students to somewhat mindlessly tweak their implementation, using a kind of
`trial-and-error' approach, without maintaining an internal justification.
Future work could investigate how students respond to feedback, for example
with a think-aloud study.

Another class of skills that may be neglected in Haskell-based exercises
arise due to the structure imposed by our interface.
In designing a test suite we usually make assumptions about
the Haskell type of the student's answer. We must then explain to the student
which type we expect as part of the exercise description.
Thus the student does not get to exercise the skill of taking a proposition
and thinking for themselves about what sort of constructive proof would be
appropriate to establish this proposition.
This gap could potentially be addressed through careful Haskell exercise
design, such as by offering students a choice of types with which to respond
using a suitable algebraic data type.
However, the most direct response may be to offer students an opportunity
to practice open-ended written proofs.

The skill of writing fluent and clear mathematical arguments
in mathematical notation is clearly not directly accessible through Haskell
exercises.
If this skill is an intended learning outcome, then perhaps written exercises
should be emphasised.
It is possible to practice communication skills by writing clear Haskell
programs as answers to Haskell exercises, but assessing these dimensions
of Haskell programs falls beyond the remit of the compiler and simple
automatic tests.
Anyway, the premise of using Haskell as a bridge to a mathematical formalism
and notation suggests that we should eventually help our students
\emph{cross} the bridge into the more conventional language of mathematics.

\subsubsection{Additional skills demanded by Haskell exercises}
\label{sec:discussion:barriers}

The main skill demanded by Haskell exercises in addition to those required to
complete a written proof is, of course, the ability to express one's ideas in
a functional programming language.
Completing Haskell-based exercises requires students to become reasonably
proficient in Haskell. 
At times, students are also required to provide some language-level 
details in the specification of the construction which would be below 
the level of abstraction appropriate for a clear written proof.
Furthermore, taking advantage of the feedback provided by the compiler
also requires the skill of interpreting Haskell error messages.

To minimise these barriers, we take care to design and select Haskell
exercises that use a minimal subset of language features, and have
answers that do not require much low-level detail. 
This sometimes involves carefully scoping problems for students to 
implement a small part of a larger program, with the aid of helper 
functions that present a clear abstraction hiding complicated parts of 
the construction (noting that adding novel helper functions to an 
exercise itself adds cognitive load for students).
The overall impact of these barriers on student outcomes in our course
requires further evaluation.

\subsection{Digital formative and summative assessment}

Rapid compiler-based and test-based feedback have the potential to enhance
student learning of the exercised skills, compared to written proof
exercises.
With traditional proof exercises, providing rapid feedback to students at
scale and on-demand is infeasible. In our case, this kind of environment can
only be provided to students during a fraction of time spent in tutor-lead
weekly tutorials
(that fraction when the tutor is available to interact directly with a small
group of students, and the syllabus allows this time to be spent on
proof-writing exercises, as opposed to some other intended learning outcome).
In comparison, Grok affords students flexible, on-demand, instant corrective
feedback from our pre-designed test suites. 
No doubt this feedback helps students notice gaps or errors in their 
specifications of constructions, and train the related skill.

Aside from providing rapid feedback in formative assessment, 
we also provide compiler-based and limited test-based
feedback to students during high-stakes tests such as final exams,
and we allows students to fix such errors and resubmit their
answers an unlimited number of times.
We believe that this policy can increase the validity of assessment by
helping students filter out small errors such as typos, syntax errors,
and well-formedness errors from their answers before they are evaluated.

When it comes to marking student responses in a
digital format compared to marking written proofs, we find that Haskell
exercises allow for more flexible and often more efficient marking
workflows compared to written proofs.
Note that as a baseline, it is always possible to print and manually mark
student programs, as if they had been written on paper in the traditional
manner.
At the other extreme, high-quality, fully automated marking of the kinds of
complex Haskell programs we are discussing in this paper would require more
sophisticated analysis programs than we typically have the resources to
commission for our course.
The key point is that, with a student's digital response, we are free to
blend manual marking work and automation in a flexible way,
and even to change this mix during marking.
We can automate tasks that are well-suited to automation, such as compilation
and testing of student programs (cf.\ a tutor poring over a student's
handwritten construction, attempting to run it through her own mental logical
compiler and correctness tests).
And we can manually handle the remainder: 
the parts that may call for human judgement, 
such as evaluating non-compiling solutions for partial credit, 
and judging student programs against a marking scheme that takes
more than test correctness into account.
Importantly, for most instance problems we can also simplify the manual 
marking residue by \emph{automatically} clustering incorrect responses.  
For example, submitted formulas or regular expressions can be grouped
by size and/or semantic equivalence, requiring a human marker to
determine the merits only of \emph{representative} solutions---something
of considerable value when enrolment numbers are large.

\section{Conclusion}
\label{sec:conclusion}

The objective behind our use of Haskell as an assessment medium in a
non-programming course has been to give computing students a bridge
to a mathematically demanding syllabus.
We saw potential benefits, both logistic and pedagogical, of
a digitisation of elements of the course~\cite{Farrugia+2022}.
Auto-marking especially is attractive, 
though it is far from clear that depth, quality and reliability
of assessment can be maintained in the transition.
We now conduct practically all assessment digitally, through Haskell.

Harnessing Haskell to do the job of traditional online question types,
such as multiple-choice, is straightforward, once students have 
completed a short intensive introduction to the language.
But we would like to also use Haskell for various open-ended
(or ``constructed response'') questions.
In this paper we have shown how we administer certain proof-style problems.
Our approach integrates seamlessly with the other question types we use---all
are delivered on one interactive learning platform
using Haskell as the assessment medium.
We have provided many examples of constructive-proof
problems that we have used in assessment, including exams.

These more sophisticated problem types are well-suited for testing
higher-order cognitive skills.
But they also tend to assume a greater mastery of Haskell,
something that is not actually expressed as an intended learning 
outcome in our course.
This raises the question of whether Haskell might present a barrier
for some students who would otherwise have no difficulty expressing
themselves in traditional mathematical language.
We have very limited data on which to base an answer,
and more research is needed.
However, the analysis presented in Section~\ref{sec:analysis}
suggests that 5--10\% of our students do find it hard to express
themselves well in Haskell when an answer calls for a combination
of Haskell features.
Given the logistic benefits of the programming approach,
we will continue to calibrate assessment tasks to find the right balance
between medium and expressiveness.

\subsection*{Acknowledgements}

We thank our teaching colleagues Anna Kalenkova, Bach Le, Billy Price, 
Rohan Hitchcock, and the rest of the COMP30026 teaching team.
We thank Kevin Kappelmann, TFPIE2022 conference attendees and our 
anonymous reviewers for helpful discussions and suggestions.

\bibliographystyle{eptcs}
\bibliography{refs}

\end{document}